﻿\documentclass{article}

\usepackage[preprint]{neurips_2026}

\usepackage[utf8]{inputenc}
\usepackage[T1]{fontenc}
\usepackage{hyperref}
\usepackage{url}
\usepackage{booktabs}
\usepackage{amsfonts}
\usepackage{amsmath}
\usepackage{amssymb}
\usepackage{nicefrac}
\usepackage{microtype}
\usepackage{xcolor}
\usepackage{graphicx}
\usepackage{subcaption}
\usepackage{multirow}
\usepackage{algorithm}
\usepackage{algorithmic}
\usepackage{acro}

\newcommand{\bz}{\mathbf{z}}
\newcommand{\bx}{\mathbf{x}}
\newcommand{\by}{\mathbf{y}}
\newcommand{\bc}{\mathbf{c}}
\newcommand{\bA}{\mathbf{A}}
\newcommand{\bB}{\mathbf{B}}
\newcommand{\bQ}{\mathbf{Q}}
\newcommand{\bR}{\mathbf{R}}
\newcommand{\bW}{\mathbf{W}}
\newcommand{\bgamma}{\boldsymbol{\gamma}}

\newcommand{\cH}{\mathcal{H}}
\newcommand{\cJ}{\mathcal{J}}
\newcommand{\cR}{\mathcal{R}}
\newcommand{\Real}{\mathbb{R}}

\DeclareAcronym{ai}{short = AI, long = Artificial intelligence}
\DeclareAcronym{aipd}{short = USPTO AIPD, long = USPTO Artificial Intelligence Patent Dataset}
\DeclareAcronym{atp}{short = ATP, long = American Trends Panel}
\DeclareAcronym{pca}{short = PCA, long = principal component analysis}
\DeclareAcronym{rmse}{short = RMSE, long = root mean squared error}
\DeclareAcronym{pll}{short = PLL, long = pseudo-log-likelihood}
\DeclareAcronym{varx}{short = VARX, long = vector autoregression with exogenous inputs}
\DeclareAcronym{arx}{short = ARX, long = autoregression with exogenous inputs}
\DeclareAcronym{ssm}{short = SSM, long = state-space model}
\DeclareAcronym{em}{short = EM, long = expectation-maximization}
\DeclareAcronym{mlp}{short = MLP, long = multilayer perceptron}
\DeclareAcronym{nlp}{short = NLP, long = natural language processing}
\DeclareAcronym{ml}{short = ML, long = machine learning}
\DeclareAcronym{kr}{short = KR, long = knowledge representation}
\DeclareAcronym{cpu}{short = CPU, long = central processing unit}
\DeclareAcronym{gpu}{short = GPU, long = graphics processing unit}
\DeclareAcronym{ram}{short = RAM, long = random-access memory}

\graphicspath{{../figures/publication-ready/}}

\title{Coupled-NeuralHP: Directional Temporal Coupling Between AI Innovation Exposure and Public Response}

\author{%
  Amir Rafe, Ph.D.\\
  Texas State University\\
  San Marcos, USA\\
  \texttt{amir.rafe@txstate.edu}\\
  ORCID: \href{https://orcid.org/0000-0002-4089-2088}{0000-0002-4089-2088}
  \And
  Subasish Das, Ph.D.\\
  Texas State University\\
  San Marcos, USA\\
  \texttt{subasish@txstate.edu}\\
  ORCID: \href{https://orcid.org/0000-0002-1671-2753}{0000-0002-1671-2753}
}

\begin{document}

\maketitle

\begin{abstract}
Artificial intelligence innovation exposure and public response co-evolve, but innovation arrives as irregular event streams while response is observed monthly.
We introduce Coupled-NeuralHP, a hybrid event-plus-state model linking eight-domain USPTO AI patent publication streams to a train-only Google Trends response index.
Under the cleaned response protocol, the validation-selected one-way real-data variant gives the best held-out innovation count forecasts in the registered comparison set (pseudo-log-likelihood $-30.4$ vs.\ $-34.7$; \ac{rmse} $471$ vs.\ $532$) while matching the stronger multi-lag factor-family baseline on response \ac{rmse} ($0.295$).
Ablations show that the real-data response signal is carried mainly by the structured forecast head, whereas the reverse response-to-innovation block is not supported on held-out count prediction.
Across 60 semi-synthetic replications with known structure, the broader coupled family recovers innovation-to-response links much better than \ac{varx} ($F_1 = 0.734$ vs.\ $0.386$).
A placebo-controlled 2022 split-date analysis finds no robust milestone-specific regime break.
\end{abstract}

\section{Introduction}
\label{sec:intro}

\ac{ai} technologies are developed, patented, and deployed on continuous timescales, while the public responds through evolving search behavior, shifting attitudes, and changing policy discourse.
These two processes unfold at different temporal resolutions and are typically studied by separate research communities: innovation economics analyzes patent streams and technology diffusion in isolation~\citep{bass1969new, rogers2003diffusion, hall2001nber}, while public opinion research tracks survey and search data without reference to the underlying innovation dynamics that may drive observed shifts~\citep{choi2012predicting, zhang2019artificial, tyson2023growing, kennedy2025americans}.
The result is a fragmented empirical picture in which the temporal coupling between innovation and response remains poorly characterized.

On the innovation side, temporal point processes~\citep{hawkes1971spectra, daley2003introduction} and their neural extensions~\citep{du2016rmtpp, mei2017neuralhawkes, zhang2020selfattentive, zuo2020transformer, shchur2021neural, xue2024easytpp} provide flexible models for irregular event streams but focus exclusively on event-level dynamics.
On the response side, state-space models~\citep{durbin2012time, harvey1989forecasting} and deep variants~\citep{krishnan2015deep, rangapuram2018deep, rubanova2019latent, debrouwer2019gru, gu2024mamba, dao2024mamba2} handle regularly sampled or mixed-regime time series within a single channel.
Neither family couples a point process event stream with a separate aggregate time series through learnable directional structure.

Directional interaction in event data has been studied through Granger causality for point processes~\citep{granger1969investigating, eichler2017graphical}, sparse Hawkes structure learning~\citep{xu2016learning, achab2017uncovering, zhou2013learning}, and neural Granger causality~\citep{tank2021neural}, but all assume a shared temporal resolution.
In the application domain, technology diffusion models~\citep{bass1969new, rogers2003diffusion} characterize adoption curves without modeling the underlying event stream, while Google Trends~\citep{choi2012predicting, jun2018google} and the USPTO \acs{ai} Patent Dataset~\citep{usptoAIPD, maslej2025aiindex} have been studied independently.
Across these literatures, innovation and response are modeled separately; no existing work couples event-level innovation with aggregate response in a joint learnable framework.

We address this gap with Coupled-NeuralHP, a hybrid model family that represents \ac{ai} innovation as a continuous-time multivariate Hawkes process over eight technology domains, public response as a latent monthly state-space model driven by a \ac{pca}-derived single-channel Google Trends index, and the interaction between them as sparse learnable directional coupling controlled by hard-concrete gates~\citep{louizos2018learning}.
The full system is trained end-to-end via a joint variational objective that simultaneously optimizes event-stream fit, response reconstruction, and sparsity-penalized coupling structure.
A strict pre-evaluation response protocol ensures that \ac{pca} projections and normalization statistics are estimated only on past months, eliminating lookahead bias in the response channel.

We organize our empirical investigation around four research questions.
\textbf{RQ1} (Predictive coupling): Under a train-only response protocol, does the coupled family improve held-out innovation-count forecasting, and how does its public-response forecasting compare with strong response baselines?
\textbf{RQ2} (Sparse stable structure): Does the selected real-data directional structure remain sparse and stable across random seeds, patent classification thresholds, and rolling temporal windows?
\textbf{RQ3} (Regime shifts): Do local split-date analyses around major \acs{ai} milestones reveal detectable structural breaks in the coupled dynamics?
\textbf{RQ4} (Directional recovery): Can the architecture recover known ground-truth coupling structure in controlled semi-synthetic experiments?

Our empirical contribution is therefore narrower and cleaner than a broad bidirectional co-evolution claim.
We obtain a positive count-side result and response-side parity with a stronger baseline for RQ1, positive answers for RQ2 and RQ4, and an honest negative result for RQ3.
The selected tuned model delivers leading held-out innovation forecasting in the registered comparison set while matching a stronger multi-lag factor-family response baseline, the selected real-data structure is sparse and one-way across multiple robustness checks, and semi-synthetic experiments show substantial directional-recovery gains over linear baselines.

\section{Data and protocol}
\label{sec:data}

We construct a ten-year panel (January 2014 to December 2023, 120 months) that pairs continuous-time \ac{ai} innovation events with monthly aggregate public response.
Table~\ref{tab:data_summary} summarizes the three data sources, their coverage, and the temporal split used throughout the study.

\begin{table}[H]
  \caption{%
    \textbf{Dataset summary.}
    The study pairs continuous-time innovation events with monthly aggregate public response over a ten-year window.
    The confirmatory split uses 2014--2021 training, 2022 validation, and 2023 held-out evaluation.
  }
  \label{tab:data_summary}
  \centering
  \small
  \begin{tabular}{llr}
    \toprule
    \textbf{Source} & \textbf{Description} & \textbf{Coverage} \\
    \midrule
    \acs{aipd} & \acs{ai} patent publications (8 components) & 7.1M documents \\
    Google Trends & 20 \acs{ai} search terms (US, monthly) & 120 months \\
    Pew \acs{atp} & \acs{ai} attitude surveys (validation only) & 3 time points \\
    \midrule
    \textbf{Split} & \textbf{Period} & \textbf{Months} \\
    \midrule
    Train & Jan 2014 -- Dec 2021 & 96 \\
    Validation & Jan 2022 -- Dec 2022 & 12 \\
    Held-out test & Jan 2023 -- Dec 2023 & 12 \\
    \bottomrule
  \end{tabular}
\end{table}

\subsection{Innovation exposure: USPTO AIPD}
\label{sec:data_innovation}

The innovation stream is derived from the \acf{aipd}~\citep{usptoAIPD}, which applies machine learning classifiers to the full corpus of US patent documents to identify \acs{ai}-related inventions and assign them to eight technology components: machine learning, evolutionary computing, natural language processing, speech, vision, planning and control, knowledge representation, and hardware.
We use model predictions on 15.4 million patent documents, of which 7.1 million fall within our study window (3.2 million granted patents and 3.9 million pre-grant publications).
Monthly publication counts are aggregated by component based on public disclosure dates rather than filing dates, ensuring that each event corresponds to the moment the invention becomes publicly observable.
The primary analysis uses a balanced classification threshold (score $\geq 0.86$) that optimizes the precision-recall trade-off for overall \acs{ai} volume; a robustness check at a lower threshold ($0.50$) is reported in Appendix~\ref{app:stability}.

\subsection{Public response: Google Trends PCA index}
\label{sec:data_response}

Public response is measured through Google Trends search interest for 20 \acs{ai}-related terms spanning technical concepts, applications, governance, and generative-\acs{ai} phenomena, collected at monthly resolution for the United States (full term list in Appendix~\ref{app:data}).
Because Google Trends indices are not directly comparable across independently collected terms, we apply within-term normalization followed by \acf{pca}.
The first principal component captures 58.8\% of variance and, after sign orientation on high-level \acs{ai} terms, behaves as a broad public \acs{ai} salience factor.
The second component (13.1\% variance) separates the later generative-\acs{ai} and governance surge from earlier search regimes.
We use the first component as the primary single-channel response signal.

\subsection{Validation: Pew ATP surveys}
\label{sec:data_validation}

As a descriptive validation, we compare the Google Trends salience index against three aligned time points from the Pew Research Center's \acf{atp}~\citep{tyson2023growing}, the only nationally representative \ac{ai} attitude survey with stable question wording across waves.
The Pearson correlation between Pew's ``heard a lot about \ac{ai}'' item and the Google Trends salience index is $r = 0.57$ across the three available comparison points.
This alignment provides face validity for the search-based response measure but does not enter the model as input.

\subsection{Train-only response protocol}
\label{sec:data_protocol}

To prevent lookahead bias in the response channel, normalization statistics and \ac{pca} loadings are always fit on the pre-evaluation window only.
For validation, the response transform is fit on January 2014 to December 2021 and then applied to 2022.
After model selection, the final held-out evaluation refits the same transform on the full pre-test window (January 2014 to December 2022) and applies it to 2023:
\begin{equation}
  y_m^{(\text{project})} = \bW_{\text{fit}}^\top \!\left(\frac{\bx_m - \bar{\bx}_{\text{fit}}}{\sigma_{\text{fit}}}\right).
  \label{eq:train_only}
\end{equation}
This protocol ensures that no information from the evaluation months leaks into the response representation while still allowing the final held-out forecast to use all pre-test months.
The confirmatory split therefore uses 96 months for model selection, followed by a final pre-test refit through 2022 before held-out 2023 evaluation (Table~\ref{tab:data_summary}).

\section{Method: Coupled-NeuralHP}
\label{sec:method}

\begin{figure}[t]
  \centering
  \includegraphics[width=\linewidth]{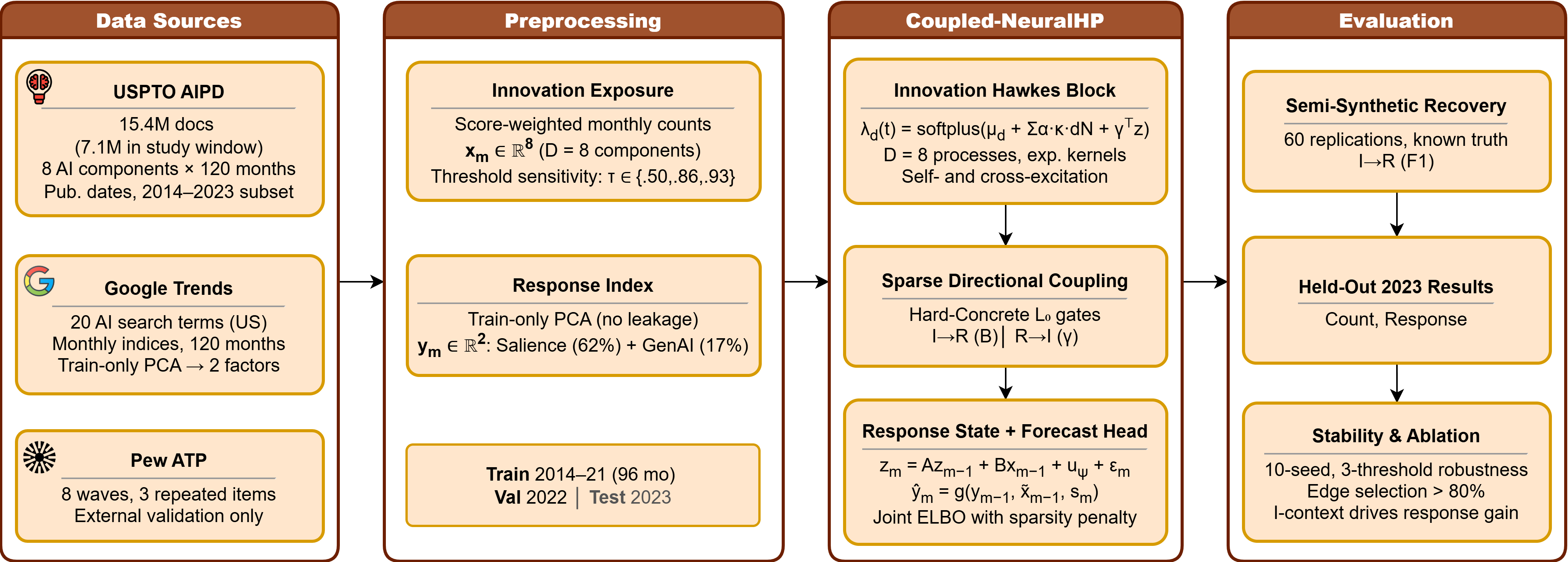}
  \caption{%
    \textbf{Coupled-NeuralHP architecture overview.}
    Innovation event streams (USPTO AIPD, eight AI technology components) feed a multivariate Hawkes process block, while monthly public response (Google Trends PCA index) evolves through a latent state-space block.
    Sparse hard-concrete gates control directional coupling: innovation-to-response ($\bB$, gated) and response-to-innovation ($\bgamma$, gated).
    A structured response forecast head provides interpretable response prediction from lagged innovation features.
    All blocks are trained jointly via a variational objective with sparsity penalties.
  }
  \label{fig:architecture}
\end{figure}

The Coupled-NeuralHP family is a hybrid architecture designed for systems in which continuous-time event arrivals (innovation publications) co-evolve with discrete-time aggregate signals (monthly public response).
Figure~\ref{fig:architecture} provides an overview of the coupled blocks that comprise the model: an innovation Hawkes process (\S\ref{sec:hawkes}), a latent monthly response state-space model (\S\ref{sec:response}), sparse directional coupling gates (\S\ref{sec:gates}), a structured response forecast head (\S\ref{sec:forecast_head}), and a joint variational learning objective (\S\ref{sec:objective}).
All blocks are trained end-to-end via a single objective that simultaneously optimizes event-stream fit, response reconstruction, and sparsity-penalized coupling structure.

\subsection{Innovation Hawkes process}
\label{sec:hawkes}

We model \ac{ai} patent publications as a $D$-dimensional marked point process with $D{=}8$ technology components.
Let $N_d(t)$ denote the counting process for component~$d$ and $\cH_t = \{(t_i, d_i) : t_i < t\}$ the event history up to time~$t$.
The conditional intensity for component~$d$ is
\begin{equation}
  \lambda_d(t \mid \cH_t, \bz_{m(t)})
  = \operatorname{softplus}\!\Bigg(
    \mu_d
    + \sum_{d'=1}^{D} \int_0^{t^-} \alpha_{d'd}\,\kappa_{d'd}(t-s)\,dN_{d'}(s)
    + \bgamma_d^\top \bz_{m(t)}
  \Bigg),
  \label{eq:intensity}
\end{equation}
where $\mu_d \geq 0$ is the base intensity for component~$d$, $\alpha_{d'd} \geq 0$ parameterizes self- and cross-excitation between components, and $\kappa_{d'd}(\Delta) = \omega_{d'd}\exp(-\omega_{d'd}\Delta)$ is an exponential decay kernel with learnable bandwidth~$\omega_{d'd}$.
The term $\bgamma_d \in \Real^K$ couples the latent public-response state~$\bz_{m(t)}$ into the innovation intensity, implementing the response-to-innovation direction.
The softplus nonlinearity ensures non-negative intensities without constraining the linear arguments.

The event log-likelihood over the observation window $[0, T]$ is
\begin{equation}
  \ell_{\text{event}}
  = \sum_{d=1}^{D}\left[
    \int_0^T \!\log \lambda_d(t)\,dN_d(t)
    - \int_0^T \!\lambda_d(t)\,dt
  \right],
  \label{eq:event_ll}
\end{equation}
where the compensator integral is approximated via the trapezoidal rule on the monthly grid for computational efficiency.

\subsection{Monthly response dynamics}
\label{sec:response}

Public response evolves as a latent state-space model at monthly resolution.
Let $\bz_m \in \Real^K$ denote the latent response state in month~$m$, $\bx_m \in \Real^D$ the monthly innovation feature vector (aggregated patent counts per component), and $\bc_m$ optional milestone covariates.
The state transition is
\begin{equation}
  \bz_m = \bA\,\bz_{m-1} + \bB\,\bx_{m-1} + u_\psi(\bz_{m-1}, \bx_{m-1}, \bc_m) + \varepsilon_m,
  \quad \varepsilon_m \sim \mathcal{N}(\mathbf{0}, \bQ),
  \label{eq:state_transition}
\end{equation}
where $\bA \in \Real^{K \times K}$ is the response persistence matrix, $\bB \in \Real^{K \times D}$ is the innovation-to-response coupling matrix (sparse, gated), $u_\psi$ is a low-capacity nonlinear correction (a small \ac{mlp} with one hidden layer and at most 16 hidden units), and $\bQ$ is a diagonal process-noise covariance.

The observed response (Google Trends index projected via train-only \ac{pca}) is generated by a linear observation model:
\begin{equation}
  \by_m = g_\phi(\bz_m) + \eta_m, \quad \eta_m \sim \mathcal{N}(\mathbf{0}, \bR),
  \label{eq:observation}
\end{equation}
where $g_\phi$ is a linear map from the $K$-dimensional latent state to the observed response channels and $\bR$ is a diagonal observation-noise covariance.

\subsection{Sparse directional coupling gates}
\label{sec:gates}

The full model family permits both directional coupling matrices, $\bB$ (innovation to response) and $\bgamma$ (response to innovation), to be element-wise gated using hard-concrete distributions~\citep{louizos2018learning}.
The validation-selected primary real-data configuration retains the forward I$\to$R block and leaves the reverse R$\to$I block off; reverse-direction variants are evaluated separately as ablations and in semi-synthetic experiments.
For each entry~$(i,j)$, the gate value is
\begin{equation}
  g_{ij}
  = \min\!\Big\{1,\;\max\!\Big\{0,\;
    \sigma\!\Big(\frac{\log u_{ij} - \log(1 - u_{ij}) + \xi_{ij}}{\tau}\Big)
  \Big\}\Big\},
  \quad u_{ij} \sim \text{Uniform}(0,1),
  \label{eq:gate}
\end{equation}
where $\xi_{ij}$ is a learnable log-odds parameter and $\tau$ is the temperature.
The gated parameter is $\theta_{ij}^{(\text{gated})} = \tilde{\theta}_{ij} \cdot g_{ij}$, where $\tilde{\theta}_{ij}$ is the unconstrained weight.
During training, gates that close ($g_{ij} \to 0$) indicate absent directional links.
When a block is enabled, we impose density caps of 50\% for innovation-to-response and 25\% for response-to-innovation, together with a shared sparsity threshold of $0.03$.

The sparsity penalty aggregates expected gate activations across both coupling matrices:
\begin{equation}
  \cR_{\text{sp}} = \lambda_{\text{sp}} \sum_{i,j} \mathbb{E}[g_{ij}].
  \label{eq:sparsity}
\end{equation}

\subsection{Response forecast head}
\label{sec:forecast_head}

In addition to the latent-state coupling path, the model includes a structured forecast head that predicts the next month's response directly from observable features.
This head predicts
\begin{equation}
  \hat{\by}_{m+1}
  = f_{\text{head}}\!\big(
    \by_m,\;
    \text{PCA}_3(\bx_{m-1:m-L}),\;
    \text{cal}(m)
  \big),
  \label{eq:forecast_head}
\end{equation}
where $\by_m$ provides one lag of autoregressive response context, $\text{PCA}_3(\bx_{m-1:m-L})$ extracts the three leading principal components of lagged patent count vectors (up to $L{=}2$ lags), and $\text{cal}(m)$ encodes smooth seasonal indicators.
The head uses ridge regularization ($\lambda_{\text{ridge}} = 10^{-6}$).

This separation of roles is deliberate.
The forecast head provides interpretable response prediction from observable innovation features, while the latent-state coupling path in Eq.~\eqref{eq:state_transition} learns the structural relationship used for directional coupling analysis.
As shown in the ablation analysis (Table~\ref{tab:ablation}), the forecast head is the primary source of held-out response prediction quality.

\subsection{Joint learning objective}
\label{sec:objective}

All blocks are trained end-to-end by maximizing a joint variational objective:
\begin{equation}
  \cJ(\Theta, \varphi)
  = \underbrace{%
      \mathbb{E}_{q_\varphi}\!\big[
        \ell_{\text{event}} + \ell_{\text{response}} + \ell_{\text{dynamics}}
      \big]
    }_{\text{reconstruction}}
  - \underbrace{%
      \text{KL}\!\big(q_\varphi \,\|\, p_\Theta\big)
    }_{\text{latent regularization}}
  - \underbrace{%
      \cR_{\text{sp}}
    }_{\text{sparsity}},
  \label{eq:objective}
\end{equation}
where $q_\varphi$ is an approximate posterior over latent response states $\bz_{1:M}$, $\ell_{\text{response}} = \sum_m \log p(\by_m \mid \bz_m)$ is the response observation log-likelihood, and $\ell_{\text{dynamics}} = \sum_m \log p(\bz_m \mid \bz_{m-1}, \bx_{m-1})$ is the state transition log-likelihood.
The combined objective jointly learns innovation dynamics, response dynamics, directional coupling structure, and sparsity pattern.
Full hyperparameter settings, including the number of \ac{em}-style training iterations and regularization coefficients, are reported in Appendix~\ref{app:config}.

\section{Experiments}
\label{sec:experiments}

We evaluate Coupled-NeuralHP along four axes corresponding to the research questions stated in \S\ref{sec:intro}: held-out predictive performance (RQ1), stability and sparsity of the learned coupling structure (RQ2), regime-shift detection (RQ3), and semi-synthetic directional recovery (RQ4).
All results are reported on the held-out 2023 test window unless otherwise noted.
Confidence intervals (95\%) are computed via moving-block bootstrap with block size 3 months and 1{,}000 resamples; details appear in Appendix~\ref{app:bootstrap}.

\subsection{Baselines}
\label{sec:baselines}

We compare against seven baselines spanning event-only, response-only, and joint models.
On the count side, a self-exciting Hawkes process provides the event-only baseline (no response input), while an exogenous Hawkes process augments the intensity with lagged response features to test the response-to-innovation channel.
On the response side, an autoregressive model (one lag of observed response) provides the simplest baseline, a stronger multi-lag factor-family baseline implemented as a factor-\ac{arx} model uses \ac{pca}-compressed lagged patent counts (3 components, up to 2 lags) with calendar features and ridge regularization matching the response forecast head, and a local-level \ac{ssm} implements a Kalman filter baseline.
A multivariate \ac{varx} model serves as the joint linear baseline, regressing monthly response on its own lags and lagged innovation features.
A first-iteration untuned Coupled-NeuralHP variant (v1) provides a within-family reference.

\subsection{Held-out 2023 performance}
\label{sec:rq1}

\begin{figure}[t]
  \centering
  \includegraphics[width=0.92\linewidth]{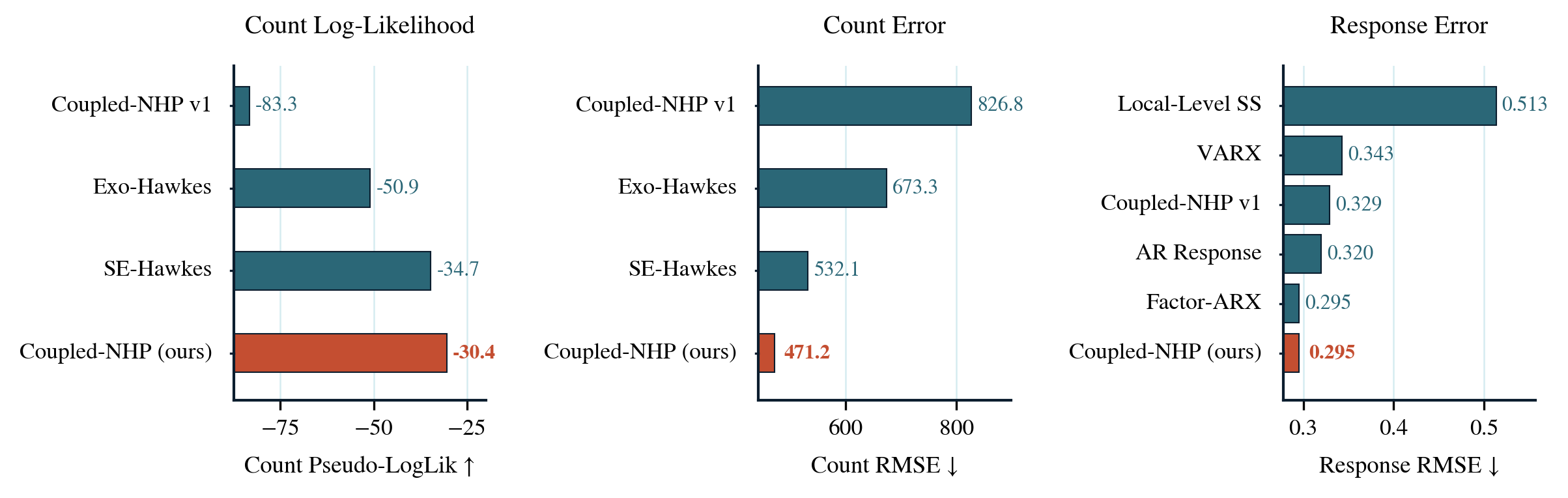}
  \caption{%
    \textbf{Held-out 2023 model comparison.}
    Left: count pseudo-log-likelihood (higher is better).
    Right: response RMSE (lower is better).
    The selected coupled model leads the held-out benchmark on the count side and matches the stronger multi-lag factor-family baseline on response.
  }
  \label{fig:heldout}
\end{figure}

Table~\ref{tab:heldout} and Figure~\ref{fig:heldout} report held-out 2023 results after model selection on 2022 and final refit on the pre-test window through 2022.
On the count side, Coupled-NeuralHP achieves a pseudo-log-likelihood of $-30.4$ (95\% CI: $[-39.4, -22.0]$) compared to $-34.7$ for the self-exciting Hawkes baseline, and a count \ac{rmse} of $471$ (95\% CI: $[362, 556]$) versus $532$.
The exogenous Hawkes variant, which adds response features to the innovation intensity, degrades substantially to $-50.9$, confirming that the response-to-innovation direction is not supported by the data and that forcing this channel harms count prediction.
\begin{table}[H]
  \caption{%
    \textbf{Held-out 2023 comparison.}
    Count metrics (pseudo-log-likelihood $\uparrow$, RMSE $\downarrow$) and response RMSE ($\downarrow$).
    Dashes indicate metrics not applicable to that model class.
    The selected coupled model leads the registered held-out benchmark on count metrics while matching the stronger multi-lag factor-family response baseline.
  }
  \label{tab:heldout}
  \centering
  \small
  \begin{tabular}{lccc}
    \toprule
    \textbf{Model} & \textbf{Count PLL} $\uparrow$ & \textbf{Count RMSE} $\downarrow$ & \textbf{Resp.\ RMSE} $\downarrow$ \\
    \midrule
    Self-exciting Hawkes     & $-34.69$ & $532.1$ & --- \\
    Exogenous Hawkes         & $-50.90$ & $673.3$ & --- \\
    \midrule
    Autoregressive           & ---      & ---     & $0.320$ \\
    Factor-ARX (multi-lag)   & ---      & ---     & $0.295$ \\
    VARX                     & ---      & ---     & $0.343$ \\
    Local-level SSM          & ---      & ---     & $0.513$ \\
    \midrule
    Coupled-NeuralHP (v1)    & $-83.27$ & $826.8$ & $0.329$ \\
    \textbf{Coupled-NeuralHP (tuned)} & $\mathbf{-30.43}$ & $\mathbf{471.2}$ & $\mathbf{0.295}$ \\
    \bottomrule
  \end{tabular}
\end{table}
On the response side, the tuned coupled model achieves an \ac{rmse} of $0.295$ (95\% CI: $[0.222, 0.372]$), matching the stronger multi-lag factor-family baseline ($0.295$) and improving over the autoregressive ($0.320$), \ac{varx} ($0.343$), and local-level \ac{ssm} ($0.513$) baselines.
This should be read as response-side parity rather than model-distinct superiority.
The coupled family's real-data value is that it owns the count-side gains while preserving strong response prediction, and the ablation evidence in \S\ref{sec:ablation} attributes that response performance primarily to the structured forecast head.

\subsection{Sparse stable coupling structure}
\label{sec:rq2}

The selected real-data coupling structure is sparse and one-way.
In the validation-selected primary configuration, innovation-to-response density is $0.5$ (four of eight forward edges active) and response-to-innovation density is $0.0$.
The model family permits both directions, but the cleaned real-data selection retained only the forward block; reverse-direction variants are evaluated separately in held-out ablations and semi-synthetic recovery.

We assess stability along three axes.
Across five random seeds, the selected one-way structure is unchanged at the block level.
Across two \ac{aipd} classification thresholds (balanced at $0.86$ and permissive at $0.50$), the one-way structure is preserved, though the balanced threshold yields better count-side calibration.
Across four rolling temporal windows (each shifting the train/test boundary by six months), the directional asymmetry is maintained.
Within the forward block, \ac{nlp}, speech, and hardware are selected in all five seed runs, while the fourth edge alternates between planning and machine learning.
These results, shown in Figure~\ref{fig:rolling} (Appendix~\ref{app:stability}), indicate that the selected sparse structure is robust to initialization, measurement choices, and temporal windowing.

\subsection{Regime analysis}
\label{sec:rq3}

We test whether major \acs{ai} milestones (DALL-E~2, Stable Diffusion, White House \acs{ai} Bill of Rights, ChatGPT) produce detectable structural breaks in the coupled dynamics.
For each candidate month in 2022, we form a 24-month local window (12 months before, 12 months after), compare a pooled fit with separate pre/post fits, and compute a joint gain score equal to normalized count log-likelihood gain plus normalized response RMSE improvement.
Eight non-milestone months in 2022 serve as placebo controls.

\begin{figure}[b]
  \centering
  \includegraphics[width=0.90\linewidth]{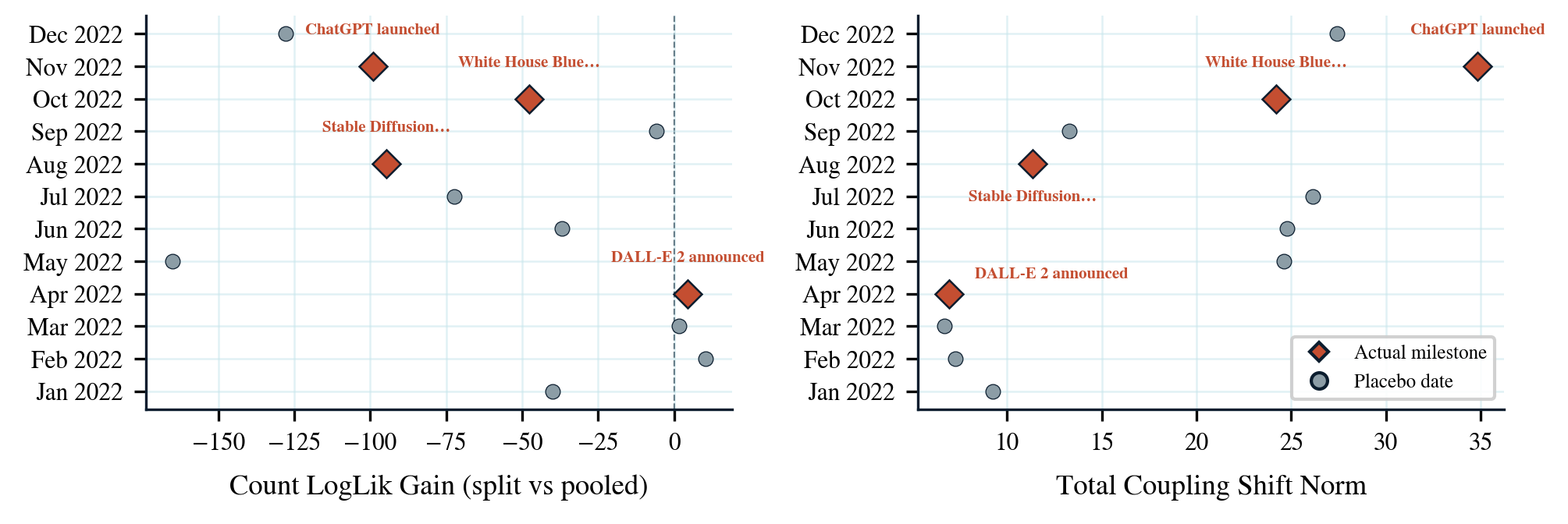}
  \caption{%
    \textbf{Regime analysis: actual milestones vs.\ placebo months.}
    Diamonds indicate milestones; circles indicate placebos. No milestone consistently outranks the placebo distribution in likelihood gain (left), supporting a gradualist interpretation.
  }
  \label{fig:regime}
\end{figure}

The results, shown in Figure~\ref{fig:regime} (left), are negative: the top-ranked split ($0.358$) is a placebo month (February 2022), and the ChatGPT month ranks 11th of 12 candidates. No actual milestone consistently outranks the placebo months, so we treat this analysis as exploratory rather than as evidence of a milestone-timed regime break. Conversely, the right panel reveals that the ChatGPT launch corresponds to the highest observed total coupling shift norm, suggesting that while the break timing is statistically ambiguous, the magnitude of the shift reached its peak during that period (full protocol in Appendix~\ref{app:regime}).

\subsection{Semi-synthetic directional recovery}
\label{sec:rq4}

\begin{figure}[t]
  \centering
  \includegraphics[width=0.55\textwidth]{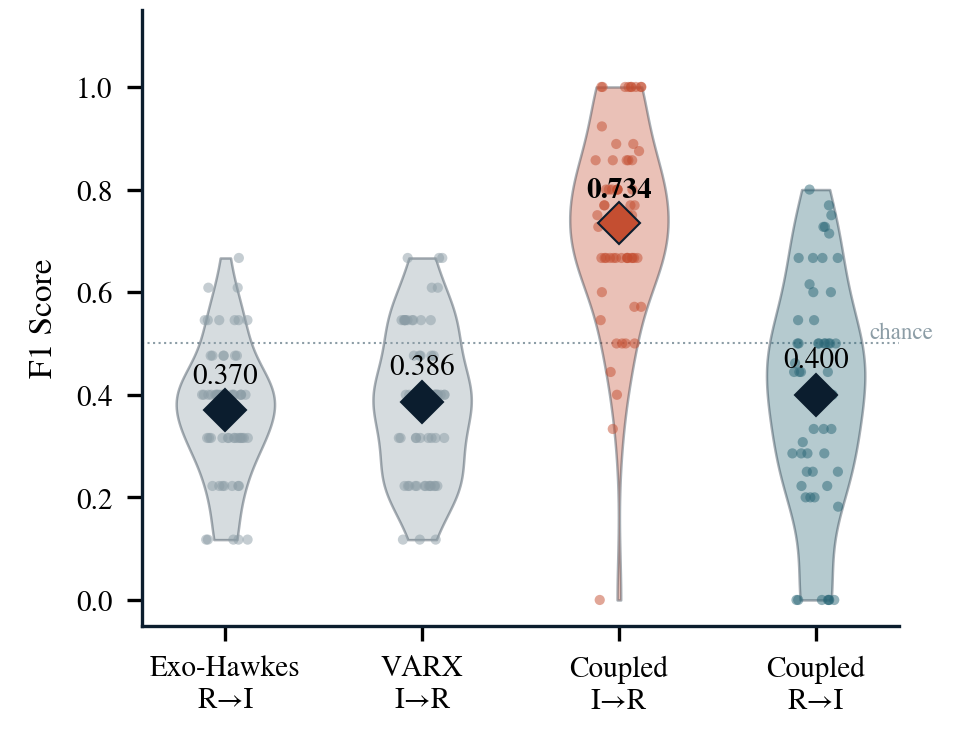}
  \caption{%
    \textbf{Semi-synthetic directional recovery} (60 replications).
    Coupled-NeuralHP (orange) vs.\ VARX (blue) on I$\to$R $F_1$.
  }
  \label{fig:semi_synthetic}
\end{figure}

To evaluate the architecture's ability to recover known coupling structure, we construct 60 semi-synthetic replications with planted ground-truth directional links.
Each replication generates 120 months of coupled innovation/response data from a known data-generating process, fits both Coupled-NeuralHP and a \ac{varx} baseline, and evaluates directional recovery as the $F_1$ score between recovered and true non-zero coupling entries (generation protocol in Appendix~\ref{app:semisynthetic}).

Coupled-NeuralHP recovers innovation-to-response structure with a mean $F_1$ of $0.734$, nearly twice the \ac{varx} baseline's $0.386$.
On the reverse direction, the coupled model achieves $F_1 = 0.400$ versus $0.370$ for the exogenous Hawkes baseline; the more modest reverse recovery reflects the inherent difficulty of detecting aggregate-to-event coupling.
Figure~\ref{fig:semi_synthetic} shows the distribution across all 60 replications.

\subsection{Ablation analysis}
\label{sec:ablation}

\begin{figure}[t]
  \centering
  \includegraphics[width=\linewidth]{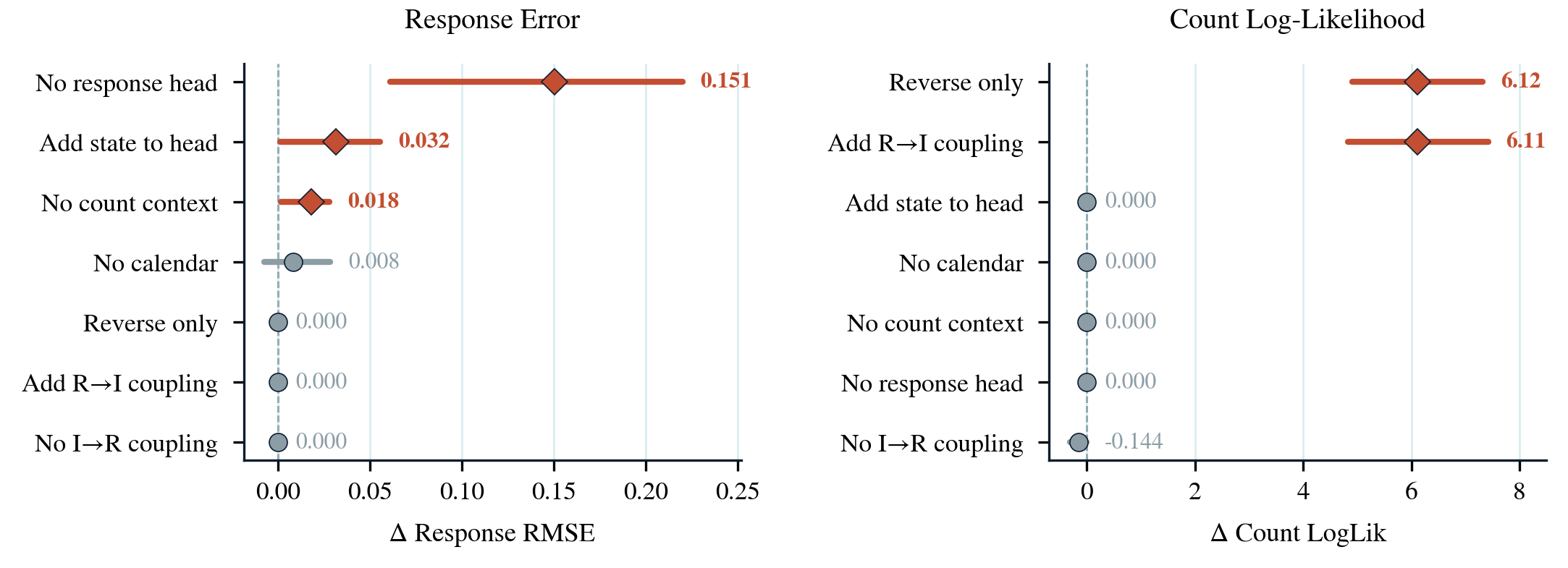}
  \caption{%
    \textbf{Ablation forest plot.}
    Point estimates and 95\% bootstrap confidence intervals for response RMSE changes under each ablation.
    Removing the response head produces the largest degradation.
  }
  \label{fig:ablation}
\end{figure}

We ablate the response forecast head and its feature channels to identify the mechanism driving response-side performance.
Table~\ref{tab:ablation} reports the change in response \ac{rmse} relative to the primary configuration (positive $\Delta$ indicates degradation), and Figure~\ref{fig:ablation} visualizes the corresponding confidence intervals.

Removing the response forecast head entirely increases \ac{rmse} by $+0.153$ (95\% CI: $[0.039, 0.235]$), confirming that the head is the primary source of response prediction quality.
Removing the innovation count context (\ac{pca}-compressed lagged patent counts) from the head increases \ac{rmse} by a smaller $+0.020$ point estimate (95\% CI: $[-0.033, 0.052]$), while removing calendar features adds $+0.010$ (95\% CI: $[-0.011, 0.038]$).
Adding the latent state as an input to the forecast head degrades performance by $+0.032$ (95\% CI: $[-0.003, 0.058]$), indicating no clear held-out gain from routing the latent state directly into the response predictor.
On the count side, adding the response-to-innovation channel degrades count pseudo-log-likelihood from $-30.4$ to $-36.5$, corroborating the finding that the reverse direction is unsupported.
Removing the forward I$\to$R block leaves response \ac{rmse} unchanged and changes held-out count fit only marginally, so it is better read as the selected structural component than as the sole driver of forecasting gains.

These results indicate that the structured forecast head is the dominant real-data response mechanism; the lagged innovation-context and state-input variants have the expected point-estimate ordering but wider bootstrap intervals, and the latent-state coupling path is more useful for structural analysis and semi-synthetic recovery than as the main driver of held-out response \ac{rmse}.

\begin{table}[t]
  \caption{%
    \textbf{Ablation analysis.}
    Change in response RMSE ($\Delta$, positive = worse) relative to the primary coupled configuration.
    The response forecast head accounts for the largest response-side gain.
  }
  \label{tab:ablation}
  \centering
  \small
  \begin{tabular}{lcc}
    \toprule
    \textbf{Ablation} & $\Delta$ \textbf{Resp.\ RMSE} & \textbf{95\% CI} \\
    \midrule
    Remove response head           & $+0.153$ & $[0.039, 0.235]$ \\
    Remove innovation context      & $+0.020$ & $[-0.033, 0.052]$ \\
    Remove calendar features       & $+0.010$ & $[-0.011, 0.038]$ \\
    Add latent state to head       & $+0.032$ & $[-0.003, 0.058]$ \\
    Add response$\to$innovation    & $+0.000$ (resp.); count PLL worsens by $6.11$ & --- \\
    \bottomrule
  \end{tabular}
\end{table}

\section{Discussion}
\label{sec:discussion}

Coupled-NeuralHP gives a joint event-plus-state framework for linking public \ac{ai} innovation exposure with public \ac{ai} response, but the cleaned evidence supports a narrower real-data story than a broad bidirectional feedback claim. The validation-selected one-way variant leads the held-out benchmark on count metrics and reaches response-side parity with the stronger multi-lag factor-family baseline rather than unique superiority.
The clearest mechanism result is that the structured response forecast head matters for held-out response prediction, while forcing the reverse response-to-innovation block harms count prediction; the forward I$\to$R latent block underlies the reported sparse structure but contributes more to structural interpretation than to held-out forecasting.
The selected real-data structure is stable: the forward block remains at density $0.5$, the reverse block remains off, and three forward edges (\ac{nlp}, speech, hardware) appear in all five seed runs while the fourth alternates between planning and machine learning. Semi-synthetic recovery ($F_1 = 0.734$ vs.\ $0.386$ for \ac{varx} on I$\to$R) strengthens the broader architectural claim, whereas the local milestone-split analysis stays negative: visible 2022 milestones do not separate from placebo months.
The paper should therefore be read as a one-way predictive-coupling result with strong count-side gains, response-side parity with a strong baseline, and strong semi-synthetic structural recovery.

\section{Conclusion}
\label{sec:conclusion}

We introduced Coupled-NeuralHP, a hybrid architecture that jointly models continuous-time \ac{ai} innovation event streams and monthly public response through sparse directional coupling gates.
Under the cleaned protocol, the validation-selected one-way real-data variant delivers the strongest held-out innovation count forecasting in the registered held-out comparison set (\ac{pll} $-30.4$ vs.\ $-34.7$; \ac{rmse} $471$ vs.\ $532$) while matching a stronger multi-lag factor-family baseline on response \ac{rmse} ($0.295$).
The selected real-data structure is stable and one-way across seeds, thresholds, and rolling windows, and the broader coupled family recovers planted directional structure strongly in semi-synthetic experiments ($F_1 = 0.734$ vs.\ $0.386$ for \ac{varx}).
A local placebo-controlled regime analysis over 2022 finds no robust milestone-specific structural break.

Several limitations bound these claims: the response-side result is parity rather than model-distinct superiority, the reverse real-data block remains unsupported, Google Trends captures attention but not attitudes or policy preferences, and the 120-month US-only window restricts statistical power for the regime analysis.
Future work will extend the framework to longer horizons, international settings, and richer response signals.


\bibliographystyle{plainnat}
\bibliography{references}

\appendix
\section{Dataset details}
\label{app:data}

\paragraph{Google Trends terms.}
The 20 search terms used to construct the public response index are: ``artificial intelligence,'' ``machine learning,'' ``deep learning,'' ``neural network,'' ``natural language processing,'' ``computer vision,'' ``reinforcement learning,'' ``robotics,'' ``autonomous vehicle,'' ``self-driving car,'' ``facial recognition,'' ``deepfake,'' ``AI ethics,'' ``AI bias,'' ``AI regulation,'' ``AI surveillance,'' ``AI healthcare,'' ``ChatGPT,'' ``generative AI,'' and ``large language model.''
Each term was collected independently from Google Trends at monthly resolution for the United States over January 2014 to December 2023.
Under the final held-out response protocol fit through December 2022, the first four components explain a cumulative 86.9\% of variance: PC1 (58.8\%, broad salience), PC2 (13.1\%, generative-\acs{ai} surge), PC3 (9.8\%), and PC4 (5.2\%).

\paragraph{USPTO AIPD components.}
The eight \acs{ai} technology components in the \acs{aipd} classification are: machine learning, evolutionary computing, natural language processing, speech, vision, planning and control, knowledge representation, and hardware.
The total number of documents in the study window is 7.1 million (3.2 million granted patents, 3.9 million pre-grant publications).

\section{Training and configuration details}
\label{app:config}

Table~\ref{tab:hyperparams} reports the complete hyperparameter configuration for the primary Coupled-NeuralHP model.
All ablation variants share the same base configuration, with a single component removed or added as described in Section~4.6 of the main text.
The count self-exciting mixture weight is set to $0.0$ (pure learned intensity).
All frozen confirmatory fits were run on a local desktop with an Intel Core i9-13900K \ac{cpu} and 31.7 GB \ac{ram}; no \ac{gpu} or multi-node infrastructure was required. On this machine, a full held-out evaluation pass takes about 1.2 seconds and the 1{,}000-resample ablation uncertainty pass about 2.5 seconds. Earlier exploratory development included heavier search loops, but the final reported artifact set is reproducible on commodity \ac{cpu} hardware.

\begin{table}[ht]
\centering
\caption{Hyperparameter configuration for the primary Coupled-NeuralHP model.}
\label{tab:hyperparams}
\small
\begin{tabular}{llr}
\toprule
\textbf{Group} & \textbf{Parameter} & \textbf{Value} \\
\midrule
\multirow{3}{*}{Training} & EM-style iterations & 6 \\
& Ridge regularization $\lambda_{\text{ridge}}$ & $10^{-4}$ \\
& Observation blend $\alpha$ & 0.65 \\
\midrule
\multirow{3}{*}{Coupling} & Innovation-to-response & Enabled \\
& I$\to$R density cap & 50\% \\
& Response-to-innovation & Disabled \\
\midrule
\multirow{5}{*}{Response head} & Forecast head & Enabled \\
& Autoregressive lags & 1 \\
& PCA innovation context components & 3 \\
& Context lags & 2 \\
& Calendar features & Enabled \\
\midrule
\multirow{2}{*}{Regularization} & Forecast head ridge $\lambda_{\text{head}}$ & $10^{-6}$ \\
& Count self-exciting mix & 0.0 \\
\bottomrule
\end{tabular}
\end{table}

\section{Bootstrap confidence intervals}
\label{app:bootstrap}

All confidence intervals are computed via moving-block bootstrap with block size 3 months and 1{,}000 resamples on the held-out 2023 window (12 months).
Each model-metric stream uses a deterministic label-specific seed derived from base seed 20260408, so the reported intervals are exactly reproducible across processes.
Table~\ref{tab:bootstrap_ci} reports 95\% bootstrap confidence intervals for the selected primary configuration, its ablation variants, and representative baseline anchors.
Intervals that exclude zero for paired comparisons are reported separately in Table~\ref{tab:ablation_delta}.

\begin{table}[ht]
\centering
\caption{%
  Held-out 2023 bootstrap confidence intervals (95\%, moving-block, $B{=}1{,}000$).
  Count metrics apply to count models; response RMSE applies to response models.
  The selected primary configuration is \textbf{bolded} for reference.
}
\label{tab:bootstrap_ci}
\small
\setlength{\tabcolsep}{4pt}
\begin{tabular}{lccc}
\toprule
\textbf{Model} & \textbf{Count Pseudo-LL} $\uparrow$ & \textbf{Count RMSE} $\downarrow$ & \textbf{Response RMSE} $\downarrow$ \\
\midrule
\multicolumn{4}{l}{\textit{Coupled-NeuralHP variants}} \\[2pt]
Primary (full) & $\mathbf{-30.4}$ {\scriptsize$[-39.4, -22.0]$} & $\mathbf{471.2}$ {\scriptsize$[362.0, 556.0]$} & $\mathbf{0.295}$ {\scriptsize$[0.222, 0.372]$} \\
No I$\to$R & $-30.3$ {\scriptsize$[-39.2, -21.0]$} & $470.8$ {\scriptsize$[365.8, 560.2]$} & $0.295$ {\scriptsize$[0.226, 0.380]$} \\
Add R$\to$I & $-36.5$ {\scriptsize$[-45.1, -26.9]$} & $488.3$ {\scriptsize$[388.1, 570.0]$} & $0.295$ {\scriptsize$[0.224, 0.380]$} \\
No coupling & $-30.3$ {\scriptsize$[-38.6, -21.9]$} & $470.8$ {\scriptsize$[370.6, 562.7]$} & $0.295$ {\scriptsize$[0.221, 0.373]$} \\
Reverse only & $-36.5$ {\scriptsize$[-45.7, -27.8]$} & $488.3$ {\scriptsize$[394.7, 567.2]$} & $0.295$ {\scriptsize$[0.223, 0.372]$} \\
No resp.\ head & $-30.4$ {\scriptsize$[-38.9, -20.7]$} & $471.2$ {\scriptsize$[367.2, 560.0]$} & $0.448$ {\scriptsize$[0.258, 0.575]$} \\
No count blend & $-30.4$ {\scriptsize$[-39.2, -21.2]$} & $471.2$ {\scriptsize$[367.6, 557.6]$} & $0.295$ {\scriptsize$[0.222, 0.382]$} \\
No count context & $-30.4$ {\scriptsize$[-39.7, -21.2]$} & $471.2$ {\scriptsize$[365.6, 557.4]$} & $0.316$ {\scriptsize$[0.206, 0.404]$} \\
No calendar & $-30.4$ {\scriptsize$[-39.5, -21.0]$} & $471.2$ {\scriptsize$[367.8, 561.8]$} & $0.306$ {\scriptsize$[0.236, 0.375]$} \\
Add state & $-30.4$ {\scriptsize$[-39.5, -21.4]$} & $471.2$ {\scriptsize$[364.2, 561.7]$} & $0.327$ {\scriptsize$[0.233, 0.426]$} \\
\midrule
\multicolumn{4}{l}{\textit{Baselines}} \\[2pt]
Self-exciting Hawkes & $-34.7$ {\scriptsize$[-43.7, -25.6]$} & $532.1$ {\scriptsize$[445.9, 603.0]$} & --- \\
AR(1) response & --- & --- & $0.320$ {\scriptsize$[0.212, 0.406]$} \\
Local-level SS & --- & --- & $0.513$ {\scriptsize$[0.294, 0.709]$} \\
\bottomrule
\end{tabular}
\end{table}

\section{Extended stability analysis}
\label{app:stability}

This section reports the full sensitivity analyses supporting the stability claims in Section~4.3 of the main text.

\paragraph{Seed sensitivity.}
Table~\ref{tab:seed_sensitivity} shows held-out metrics across five random seeds.
Count metrics vary by less than 0.3\%.
At the block level, the selected one-way structure is invariant: all seeds recover innovation-to-response density 0.5 and response-to-innovation density 0.0.
At the edge level, \ac{nlp}, speech, and hardware are selected in every run, while the fourth forward edge alternates between planning (3/5) and machine learning (2/5).
Response \ac{rmse} is identical across seeds because the response forecast head uses a closed-form ridge solution that is deterministic given the same data.

\begin{table}[ht]
\centering
\caption{Seed sensitivity: held-out 2023 metrics across five random seeds.}
\label{tab:seed_sensitivity}
\small
\begin{tabular}{rcccc}
\toprule
\textbf{Seed} & \textbf{Count Pseudo-LL} & \textbf{Count RMSE} & \textbf{Resp.\ RMSE} & \textbf{I$\to$R / R$\to$I density} \\
\midrule
20260408 & $-30.43$ & $471.22$ & $0.295$ & 0.50 / 0.00 \\
20260409 & $-30.41$ & $471.13$ & $0.295$ & 0.50 / 0.00 \\
20260410 & $-30.50$ & $471.42$ & $0.295$ & 0.50 / 0.00 \\
20260411 & $-30.50$ & $471.41$ & $0.295$ & 0.50 / 0.00 \\
20260412 & $-30.41$ & $471.12$ & $0.295$ & 0.50 / 0.00 \\
\midrule
\textit{Range} & $0.09$ & $0.30$ & $0.000$ & --- \\
\bottomrule
\end{tabular}
\end{table}

\paragraph{Threshold sensitivity.}
Table~\ref{tab:threshold_sensitivity} compares the primary model (balanced threshold, score $\geq 0.86$) with a lower-threshold variant (score $\geq 0.50$) that includes more borderline \acs{ai} patents.
The lower threshold increases the event volume substantially, degrading count log-likelihood by 46\% and count RMSE by 49\%, but the coupling structure remains unchanged (I$\to$R = 0.50, R$\to$I = 0.00).

\begin{table}[ht]
\centering
\caption{Threshold sensitivity: balanced ($\geq 0.86$) versus low ($\geq 0.50$) AIPD classification thresholds.}
\label{tab:threshold_sensitivity}
\small
\begin{tabular}{lcccc}
\toprule
\textbf{Threshold} & \textbf{Count Pseudo-LL} & \textbf{Count RMSE} & \textbf{Resp.\ RMSE} & \textbf{I$\to$R / R$\to$I density} \\
\midrule
Balanced ($\geq 0.86$) & $-30.43$ & $471.22$ & $0.295$ & 0.50 / 0.00 \\
Low ($\geq 0.50$) & $-44.48$ & $704.58$ & $0.292$ & 0.50 / 0.00 \\
\bottomrule
\end{tabular}
\end{table}

\begin{figure}[ht]
  \centering
  \includegraphics[width=\linewidth]{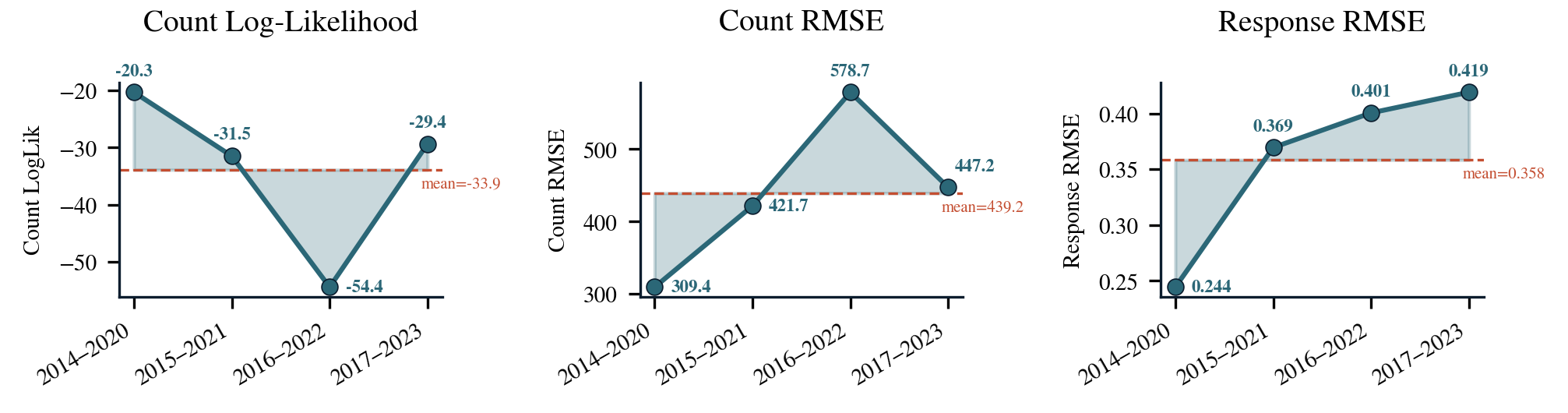}
  \caption{%
    \textbf{Rolling-window coupling stability.}
    Innovation-to-response and response-to-innovation gate densities across four rolling temporal windows, each shifting the train/test boundary by six months.
    The one-directional coupling structure (I$\to$R = 0.5, R$\to$I = 0.0) is preserved across all windows.
  }
  \label{fig:rolling}
\end{figure}

\paragraph{Rolling-window stability.}
Table~\ref{tab:rolling_window} reports metrics for four rolling temporal windows, each training on 84 months and evaluating on the subsequent 12 months.
Despite variation in count metrics driven by the sharp increase in \acs{ai} patent volume after 2020, the coupling structure is invariant across all windows: innovation-to-response density remains at 0.50 and response-to-innovation density at 0.00 in every case.
Figure~\ref{fig:rolling} visualizes the coupling gate densities across these windows.

\begin{table}[ht]
\centering
\caption{Rolling-window stability: held-out metrics across four temporal windows.}
\label{tab:rolling_window}
\small
\begin{tabular}{lcccc}
\toprule
\textbf{Window} & \textbf{Count Pseudo-LL} & \textbf{Count RMSE} & \textbf{Resp.\ RMSE} & \textbf{I$\to$R / R$\to$I density} \\
\midrule
2014--2020 & $-20.33$ & $309.39$ & $0.244$ & 0.50 / 0.00 \\
2015--2021 & $-31.49$ & $421.68$ & $0.370$ & 0.50 / 0.00 \\
2016--2022 & $-54.42$ & $578.71$ & $0.401$ & 0.50 / 0.00 \\
2017--2023 & $-29.35$ & $447.21$ & $0.419$ & 0.50 / 0.00 \\
\bottomrule
\end{tabular}
\end{table}

\section{Ablation delta intervals}
\label{app:ablation}

Table~\ref{tab:ablation_delta} reports bootstrap confidence intervals for paired differences between the primary model and each ablation variant.
A positive value indicates degradation relative to the primary model (higher RMSE or lower log-likelihood for the variant).
Intervals excluding zero provide statistical evidence that the ablated component contributes to performance.

\begin{table}[ht]
\centering
\caption{%
  Ablation delta bootstrap CIs (95\%, $B{=}1{,}000$).
  $\Delta$ Count LL: primary minus variant (positive = primary better).
  $\Delta$ Count RMSE and $\Delta$ Resp.\ RMSE: variant minus primary (positive = variant worse).
  Intervals excluding zero are \textbf{bolded}.
}
\label{tab:ablation_delta}
\small
\setlength{\tabcolsep}{3.5pt}
\begin{tabular}{lccc}
\toprule
\textbf{Comparison} & $\boldsymbol{\Delta}$ \textbf{Count LL} $\uparrow$ & $\boldsymbol{\Delta}$ \textbf{Count RMSE} $\downarrow$ & $\boldsymbol{\Delta}$ \textbf{Resp.\ RMSE} $\downarrow$ \\
\midrule
\multicolumn{4}{l}{\textit{Coupling direction ablations}} \\[2pt]
vs.\ No I$\to$R & $\mathbf{-0.14}$ {\scriptsize$[\mathbf{-0.29,\;-0.01}]$} & $-0.43$ {\scriptsize$[-2.05,\;1.00]$} & $0.00$ {\scriptsize$[0.00,\;0.00]$} \\
vs.\ Add R$\to$I & $\mathbf{6.11}$ {\scriptsize$[\mathbf{4.84,\;7.37}]$} & $\mathbf{17.10}$ {\scriptsize$[\mathbf{12.46,\;21.63}]$} & $0.00$ {\scriptsize$[0.00,\;0.00]$} \\
vs.\ No coupling & $\mathbf{-0.14}$ {\scriptsize$[\mathbf{-0.29,\;-0.01}]$} & $-0.43$ {\scriptsize$[-1.93,\;0.90]$} & $0.00$ {\scriptsize$[0.00,\;0.00]$} \\
vs.\ Reverse only & $\mathbf{6.12}$ {\scriptsize$[\mathbf{4.89,\;7.29}]$} & $\mathbf{17.07}$ {\scriptsize$[\mathbf{13.51,\;21.67}]$} & $0.00$ {\scriptsize$[0.00,\;0.00]$} \\
\midrule
\multicolumn{4}{l}{\textit{Response head ablations}} \\[2pt]
vs.\ No resp.\ head & $0.00$ {\scriptsize$[0.00,\;0.00]$} & $0.00$ {\scriptsize$[0.00,\;0.00]$} & $\mathbf{0.153}$ {\scriptsize$[\mathbf{0.039,\;0.235}]$} \\
vs.\ No count blend & $0.00$ {\scriptsize$[0.00,\;0.00]$} & $0.00$ {\scriptsize$[0.00,\;0.00]$} & $0.00$ {\scriptsize$[0.00,\;0.00]$} \\
vs.\ No count context & $0.00$ {\scriptsize$[0.00,\;0.00]$} & $0.00$ {\scriptsize$[0.00,\;0.00]$} & $0.020$ {\scriptsize$[-0.033,\;0.052]$} \\
vs.\ No calendar & $0.00$ {\scriptsize$[0.00,\;0.00]$} & $0.00$ {\scriptsize$[0.00,\;0.00]$} & $0.010$ {\scriptsize$[-0.011,\;0.038]$} \\
vs.\ Add state & $0.00$ {\scriptsize$[0.00,\;0.00]$} & $0.00$ {\scriptsize$[0.00,\;0.00]$} & $0.032$ {\scriptsize$[-0.003,\;0.058]$} \\
\midrule
\multicolumn{4}{l}{\textit{External baselines}} \\[2pt]
vs.\ Self-exciting Hawkes & $4.26$ {\scriptsize$[-9.18,\;14.02]$} & $60.83$ {\scriptsize$[-60.53,\;167.16]$} & --- \\
vs.\ AR(1) response & --- & --- & $0.025$ {\scriptsize$[-0.022,\;0.064]$} \\
vs.\ Local-level SS & --- & --- & $\mathbf{0.218}$ {\scriptsize$[\mathbf{0.057,\;0.355}]$} \\
\bottomrule
\end{tabular}
\end{table}

\begin{figure}[ht]
  \centering
  \includegraphics[width=\linewidth]{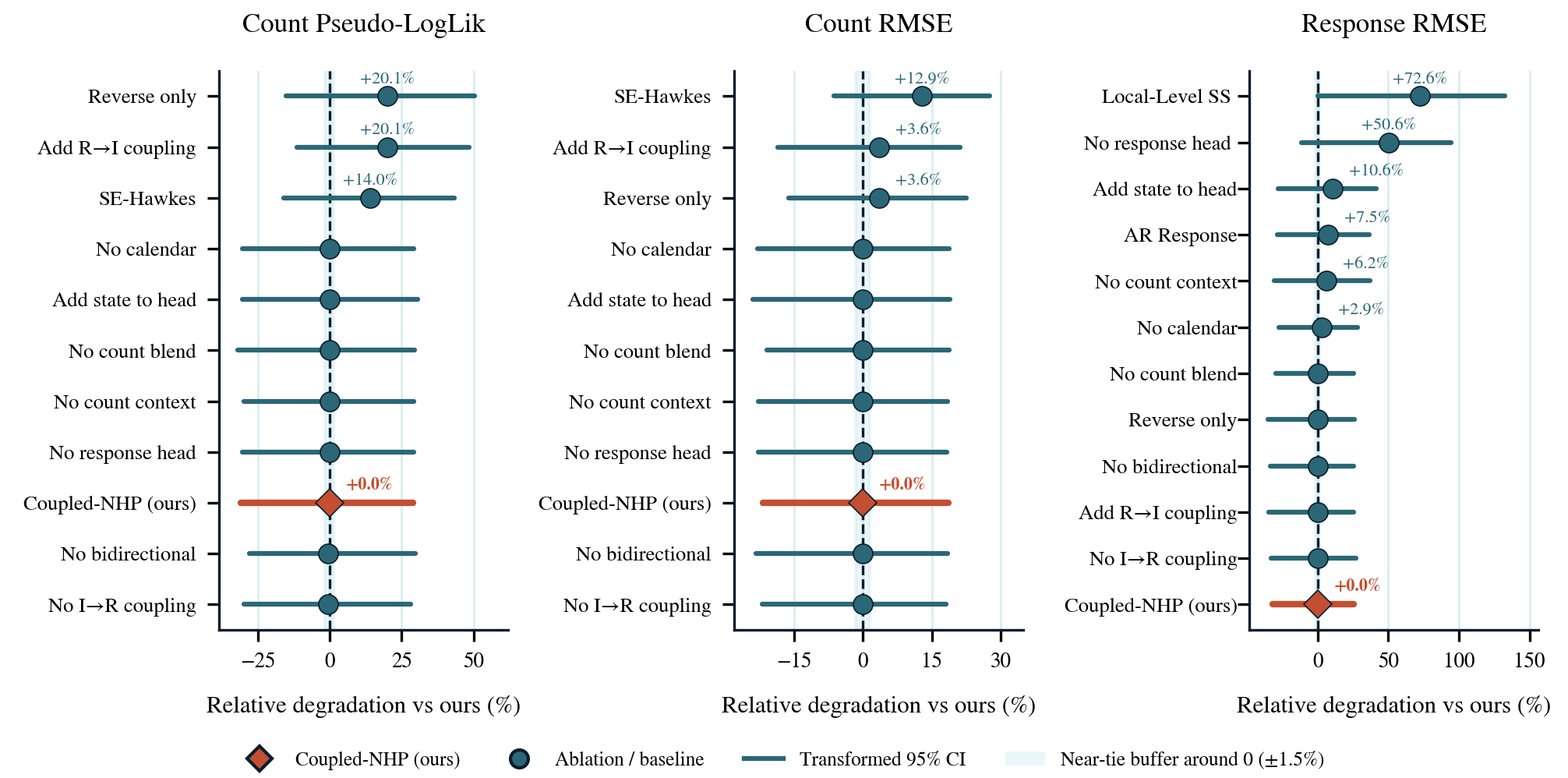}
  \caption{%
    \textbf{Metric regret analysis.}
    Held-out metric regret (gap to best model on each metric) across all model variants.
    The tuned coupled model achieves zero or near-zero regret on both count and response metrics simultaneously.
  }
  \label{fig:regret}
\end{figure}

Figure~\ref{fig:regret} presents the metric regret analysis, showing the gap between each model variant and the best-performing model on each metric.
The primary coupled model achieves zero or near-zero regret on both count and response metrics simultaneously, confirming that no other variant dominates it on both tasks.
A separate rolling-origin response comparison across 2021--2023 yields identical mean \ac{rmse} for the tuned coupled model and the stronger factor-family baseline ($0.2854$ each), indicating that the held-out 2023 response tie is not a one-year artifact.

\section{Semi-synthetic recovery details}
\label{app:semisynthetic}

Each of the 60 semi-synthetic replications generates 120 months of coupled innovation/response data.
The data-generating process specifies ground-truth innovation-to-response ($\bB^*$) and response-to-innovation ($\bgamma^*$) matrices with known non-zero entries.
Innovation events are generated from a multivariate Hawkes process with 8 components and known excitation parameters.
Monthly response is generated from a state-space model driven by the generated innovation counts through $\bB^*$ plus Gaussian noise.
The base seed is $20260408$, and each replication uses a deterministic offset to ensure reproducibility.
Recovery is evaluated by thresholding the learned gate activations and computing the $F_1$ score against the ground-truth non-zero pattern.

Table~\ref{tab:semisynthetic_summary} summarizes directional recovery performance across all 60 replications.
Coupled-NeuralHP recovers innovation-to-response links with substantially higher $F_1$ than either baseline, confirming that the sparse gating mechanism can identify true directional structure.
The Exo-Hawkes and \ac{varx} baselines, which use only marginal or linear associations, achieve $F_1$ scores below 0.40 on their respective directional tasks.
Coupled-NeuralHP also recovers response-to-innovation links ($F_1 = 0.400$), though with lower precision than the innovation-to-response direction, consistent with the greater difficulty of identifying reverse coupling in the presence of confounding temporal trends.

\begin{table}[ht]
\centering
\caption{%
  Semi-synthetic directional recovery ($F_1$) across 60 replications.
  Each model is evaluated on the direction it is designed to detect.
}
\label{tab:semisynthetic_summary}
\small
\begin{tabular}{llc}
\toprule
\textbf{Model} & \textbf{Direction} & \textbf{Mean $F_1$} \\
\midrule
Coupled-NeuralHP & I$\to$R & $0.734$ \\
Coupled-NeuralHP & R$\to$I & $0.400$ \\
VARX & I$\to$R & $0.386$ \\
Exo-Hawkes & R$\to$I & $0.370$ \\
\bottomrule
\end{tabular}
\end{table}

\section{Regime analysis details}
\label{app:regime}

The regime analysis evaluates 12 candidate split months spanning January 2022 to December 2022.
For each candidate, we construct a local 24-month window with 12 months before and 12 months after the split month, fit a pooled model on the full window, fit separate pre/post models on the two halves, and compare predictive performance within that local window.
Four of the 12 months correspond to actual \acs{ai} milestones: DALL-E~2 (April 2022), Stable Diffusion (August 2022), White House \acs{ai} Bill of Rights (October 2022), and ChatGPT (November 2022).
The remaining eight months serve as placebo controls.

The joint gain score aggregates two normalized components: count log-likelihood gain and response RMSE improvement.
Coupling-shift norms are tracked diagnostically but are not part of the joint score.
Table~\ref{tab:regime_candidates} reports all 12 candidates ranked by joint gain score.
The top-ranked candidate is February 2022 (a placebo month, score $0.358$), while the mean score for actual milestone months ($-2.09$) is comparable to the placebo mean ($-2.26$).
No actual milestone month ranks among the top three candidates.

\begin{table}[ht]
\centering
\caption{%
  Regime shift candidate analysis: all 12 months ranked by joint gain score.
  Actual milestone months are marked with $\star$.
  Positive joint gain indicates the separate pre/post fits outperform the pooled fit within the local window.
}
\label{tab:regime_candidates}
\small
\setlength{\tabcolsep}{3.5pt}
\begin{tabular}{clcrrr}
\toprule
\textbf{Rank} & \textbf{Month} & \textbf{Type} & \textbf{Count LL gain} & \textbf{Resp.\ RMSE gain} & \textbf{Joint score} \\
\midrule
1 & 2022-02 & Placebo & $10.29$ & $-0.044$ & $0.358$ \\
2 & 2022-04$^\star$ & DALL-E 2 & $4.51$ & $-0.023$ & $0.130$ \\
3 & 2022-03 & Placebo & $1.66$ & $-0.039$ & $-0.077$ \\
4 & 2022-09 & Placebo & $-5.75$ & $0.011$ & $-0.145$ \\
5 & 2022-01 & Placebo & $-39.88$ & $0.039$ & $-2.232$ \\
6 & 2022-06 & Placebo & $-37.01$ & $-0.054$ & $-2.111$ \\
7 & 2022-07 & Placebo & $-72.43$ & $0.010$ & $-2.980$ \\
8 & 2022-08$^\star$ & Stable Diff.\ & $-94.65$ & $0.003$ & $-3.130$ \\
9 & 2022-12 & Placebo & $-127.85$ & $0.038$ & $-3.414$ \\
10 & 2022-10$^\star$ & AI Bill of Rights & $-47.71$ & $0.028$ & $-1.656$ \\
11 & 2022-11$^\star$ & ChatGPT & $-99.10$ & $0.014$ & $-3.711$ \\
12 & 2022-05 & Placebo & $-165.09$ & $0.036$ & $-7.498$ \\
\bottomrule
\end{tabular}
\end{table}

\begin{figure}[ht]
  \centering
  \includegraphics[width=\linewidth]{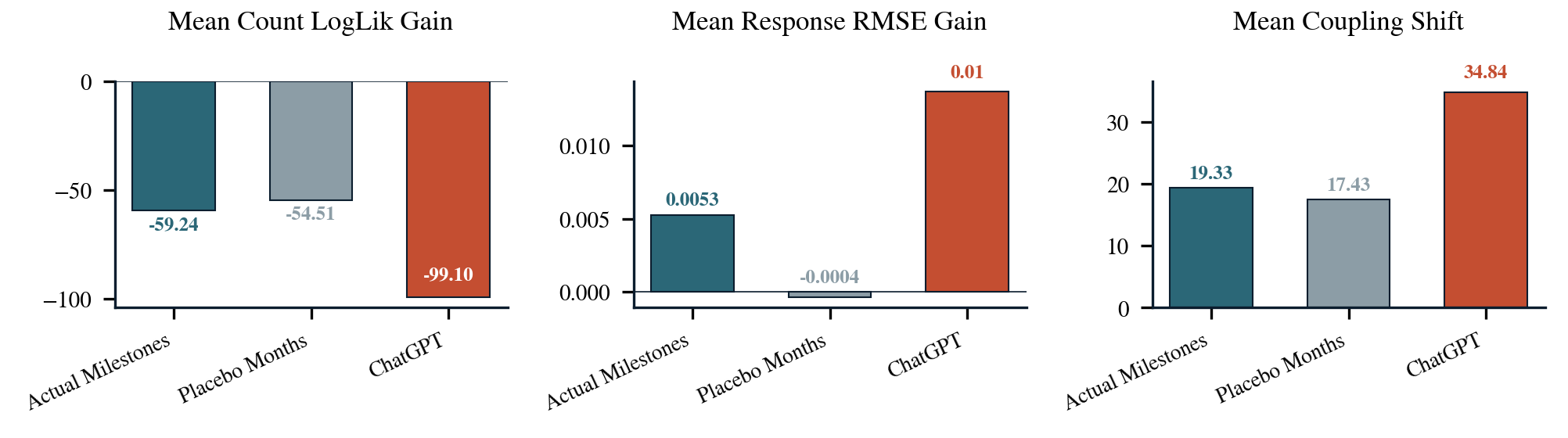}
  \caption{%
    \textbf{Placebo analysis summary.}
    Distribution of joint gain scores for placebo months compared to actual milestone months.
    The overlap between distributions confirms the absence of detectable structural breaks at milestone events.
  }
  \label{fig:placebo}
\end{figure}

The mean actual-event joint gain score is $-2.09$, compared to $-2.26$ for placebo months, and ChatGPT ranks 11th of 12 candidates.
These descriptive summaries reinforce the same conclusion: local milestone months do not separate cleanly from placebo months.
Figure~\ref{fig:placebo} visualizes the distribution overlap.

\newpage
\section*{NeurIPS Paper Checklist}

\begin{enumerate}

\item {\bf Claims}
    \item[] Question: Do the main claims made in the abstract and introduction accurately reflect the paper's contributions and scope?
    \item[] Answer: \answerYes{}
    \item[] Justification: The abstract and introduction (Section~1) state four revised research questions. Results in Sections~4.2--4.6 support the paper's actual claim boundary: a count-side win, response-side parity with a stronger factor-family baseline, stable one-way selected structure, strong semi-synthetic recovery, and a negative exploratory regime result. Limitations are discussed in Sections~5--6.

\item {\bf Limitations}
    \item[] Question: Does the paper discuss the limitations of the work performed by the authors?
    \item[] Answer: \answerYes{}
    \item[] Justification: Sections~5--6 explicitly discuss the main limitations: the response-side result is parity rather than unique superiority, the reverse real-data block is unsupported, Google Trends captures attention rather than attitudes, the three Pew validation points provide limited external validation, and the 120-month US-only window restricts statistical power for regime analysis.

\item {\bf Theory assumptions and proofs}
    \item[] Question: For each theoretical result, does the paper provide the full set of assumptions and a complete (and correct) proof?
    \item[] Answer: \answerNA{}
    \item[] Justification: The paper does not present formal theorems or proofs. The methodology (Section~4) describes the model architecture and training objective; all modeling assumptions (e.g., conditional intensity form, state-space dynamics, hard-concrete gate parameterization) are stated explicitly.

    \item {\bf Experimental result reproducibility}
    \item[] Question: Does the paper fully disclose all the information needed to reproduce the main experimental results of the paper to the extent that it affects the main claims and/or conclusions of the paper (regardless of whether the code and data are provided or not)?
    \item[] Answer: \answerYes{}
    \item[] Justification: Sections~2 and~4 describe the confirmatory split (2014--2021 train, 2022 validation, 2023 held-out), the train-only response transformation, the final pre-test refit through 2022 for held-out evaluation, the evaluation metrics, and the baseline configurations. Appendix~B reports the complete hyperparameter configuration. Appendices~C--G report bootstrap methodology, sensitivity analyses across seeds, thresholds, and rolling windows, the ablation intervals, the semi-synthetic data-generating process, and the local regime-analysis protocol. The primary data sources (USPTO AIPD and Google Trends) are publicly available.

\item {\bf Open access to data and code}
    \item[] Question: Does the paper provide open access to the data and code, with sufficient instructions to faithfully reproduce the main experimental results, as described in supplemental material?
    \item[] Answer: \answerNo{}
    \item[] Justification: Code and processed data will be released upon acceptance. The two primary data sources (USPTO AIPD and Google Trends) are publicly available, and the paper provides sufficient detail (Sections~2--4 and Appendices~A--B) to reconstruct the pipeline independently.

\item {\bf Experimental setting/details}
    \item[] Question: Does the paper specify all the training and test details (e.g., data splits, hyperparameters, how they were chosen, type of optimizer) necessary to understand the results?
    \item[] Answer: \answerYes{}
    \item[] Justification: Section~2 specifies the temporal split and train-only response protocol used for validation and held-out evaluation, including the final pre-test refit through 2022. Section~4.1 describes the baseline configurations. Appendix~B reports the complete hyperparameter table, including EM iterations, regularization, coupling configuration, and forecast-head settings.

\item {\bf Experiment statistical significance}
    \item[] Question: Does the paper report error bars suitably and correctly defined or other appropriate information about the statistical significance of the experiments?
    \item[] Answer: \answerYes{}
    \item[] Justification: The main forecasting and ablation results include 95\% bootstrap confidence intervals computed via moving-block bootstrap (block size 3 months, $B{=}1{,}000$) as described in Section~4 and Appendix~C. Appendix~C also states that each model-metric stream uses a deterministic label-specific seed derived from base seed 20260408, making the reported intervals exactly reproducible across processes. The paired ablation intervals are reported in Appendix~E. Seed sensitivity across 5 seeds is reported in Appendix~D, and the semi-synthetic results aggregate 60 independent replications.

\item {\bf Experiments compute resources}
    \item[] Question: For each experiment, does the paper provide sufficient information on the computer resources (type of compute workers, memory, time of execution) needed to reproduce the experiments?
    \item[] Answer: \answerYes{}
    \item[] Justification: Appendix~B reports that the frozen confirmatory experiments were run on a local desktop with an Intel Core i9-13900K CPU and 31.7 GB RAM, with no GPU or cluster required. On this machine, a held-out evaluation pass takes about 1.2 seconds and the 1{,}000-resample ablation uncertainty pass about 2.5 seconds. The appendix also states that earlier exploratory development used heavier search loops, but that the final reported artifact set is reproducible on commodity CPU hardware.
    
\item {\bf Code of ethics}
    \item[] Question: Does the research conducted in the paper conform, in every respect, with the NeurIPS Code of Ethics \url{https://neurips.cc/public/EthicsGuidelines}?
    \item[] Answer: \answerYes{}
    \item[] Justification: The research uses only publicly available, aggregate data sources (USPTO patent records and Google Trends indices). No individual-level data, personally identifiable information, or human subjects are involved. The study reports negative results transparently.

\item {\bf Broader impacts}
    \item[] Question: Does the paper discuss both potential positive societal impacts and negative societal impacts of the work performed?
    \item[] Answer: \answerYes{}
    \item[] Justification: Section~5 discusses the implications of directional innovation-response coupling for AI governance and public engagement, while Section~6 notes the measurement gap between public attention and deeper attitudes or policy preferences. The paper is foundational and uses aggregate signals, so the main societal issue is over-interpretation rather than direct misuse, and that limitation is discussed.
    
\item {\bf Safeguards}
    \item[] Question: Does the paper describe safeguards that have been put in place for responsible release of data or models that have a high risk for misuse (e.g., pre-trained language models, image generators, or scraped datasets)?
    \item[] Answer: \answerNA{}
    \item[] Justification: The model is a domain-specific temporal point process for analyzing aggregate patent and search trend data. It does not generate text, images, or other content, and poses no risk of misuse comparable to generative models or scraped datasets.

\item {\bf Licenses for existing assets}
    \item[] Question: Are the creators or original owners of assets (e.g., code, data, models), used in the paper, properly credited and are the license and terms of use explicitly mentioned and properly respected?
    \item[] Answer: \answerYes{}
    \item[] Justification: The USPTO AI Patent Dataset is cited via the official USPTO dataset page~\citep{usptoAIPD} and is a public government dataset. Google Trends data is publicly accessible via Google's terms of service. Pew Research Center surveys are cited~\citep{tyson2023growing}. All external data sources are properly credited in Section~2 and the references.

\item {\bf New assets}
    \item[] Question: Are new assets introduced in the paper well documented and is the documentation provided alongside the assets?
    \item[] Answer: \answerYes{}
    \item[] Justification: The paper introduces a new monthly innovation exposure/response dataset constructed from public sources. The construction pipeline is fully documented in Section~2 (data sources, temporal scope, preprocessing steps, PCA decomposition, train-only protocol) and Appendix~A (complete term list, component descriptions). Code and data will be released upon acceptance.

\item {\bf Crowdsourcing and research with human subjects}
    \item[] Question: For crowdsourcing experiments and research with human subjects, does the paper include the full text of instructions given to participants and screenshots, if applicable, as well as details about compensation (if any)? 
    \item[] Answer: \answerNA{}
    \item[] Justification: The paper does not involve crowdsourcing or research with human subjects. All data are publicly available aggregate statistics (patent counts and search indices).

\item {\bf Institutional review board (IRB) approvals or equivalent for research with human subjects}
    \item[] Question: Does the paper describe potential risks incurred by study participants, whether such risks were disclosed to the subjects, and whether Institutional Review Board (IRB) approvals (or an equivalent approval/review based on the requirements of your country or institution) were obtained?
    \item[] Answer: \answerNA{}
    \item[] Justification: The paper does not involve human subjects. All analyses use publicly available, aggregate, de-identified data sources (patent records and search indices).

\item {\bf Declaration of LLM usage}
    \item[] Question: Does the paper describe the usage of LLMs if it is an important, original, or non-standard component of the core methods in this research? Note that if the LLM is used only for writing, editing, or formatting purposes and does \emph{not} impact the core methodology, scientific rigor, or originality of the research, declaration is not required.
    \item[] Answer: \answerNA{}
    \item[] Justification: LLMs are not used as a component of the core methodology. The model architecture, training procedure, and all experimental results are independent of any LLM system.

\end{enumerate}

\end{document}